\documentclass[10pt]{wlscirep}
\usepackage[utf8]{inputenc}
\usepackage[T1]{fontenc}
\usepackage[ruled,vlined]{algorithm2e}
\usepackage{cite}
\usepackage{tabularx}
\title{High-throughput GPU layered decoder of multi-edge type low density parity check codes in continuous-variable quantum
    key distribution systems}

\author[1]{Yang~Li}
\author[2]{Xiaofang~Zhang}
\author[3,*]{Yong~Li}
\author[1,*]{Bingjie~Xu}
\author[1]{Li~Ma}
\author[1]{Jie~Yang}
\author[1]{Wei~Huang}
\affil[1]{Institute of Southwestern Communication, Science and
Technology on Security Communication Laboratory,  Chengdu 610041,
China.} \affil[2]{Chongqing University of Posts and
Telecommunications, School of Communication and Information
Engineering, Chongqing, 400065, China.} \affil[3]{Chongqing
University, College of Computer Science, Chongqing, 400044,
China.}

\affil[*]{yongli@cqu.edu.cn, xbjpku@pku.edu.cn}


\begin{abstract}
The decoding throughput in the postprocessing is one of the
bottlenecks for  a continuous-variable quantum key distribution
(CV-QKD) system. In this paper, we propose a layered decoder to
decode quasi-cyclic multi-edge type LDPC (QC-MET-LDPC) codes based
on graphic processing unit (GPU) in continuous-variable quantum
key distribution (CV-QKD) systems. We optimize the storage method
of the parity check matrix, merge the sub-matrices which are
unrelated, and decode multiple codewords in parallel on GPU.
Simulation results demonstrate that the average decoding speed of
LDPC codes with three typical code rates, i.e., 0.1, 0.05 and
0.02, is up to 64.11Mbits/s, 48.65Mbits/s and 39.51Mbits/s,
respectively, when decoding 128 codewords of length $10^6$
simultaneously without early termination.
\end{abstract}
\begin{document}

\flushbottom
\maketitle
%
%
\thispagestyle{empty}

\section*{Introduction}

In the current data-driven era, the demand of
secret communications has been increasingly desired. A
straightforward approach to secret communications is achieved by
encryption techniques. However, encryption with the key, which is
established with algorithms based on the assumptions of
computation complexity, may no longer guarantee data security, as
the large-scale universal quantum computer being reachable.

Quantum key distribution (QKD), which establishes a secret key
between two remote participants based on the quantum mechanics
principles, can provide the guaranteed security between two
participants by using a one-time-pad encryption algorithm to
encrypt and decrypt data\cite{inproceedings,Diamanti2016}. So far,
QKD has been quickly developed in both theory and experiment since
the groundbreaking work of Bennett and Brassard in
1984\cite{inproceedings}, and become one of the most mature
branches of quantum information technologies.

There exist two categories of QKD. One is the discrete variable QKD (DV-QKD) where the key information is
encoded on discrete Hilbert space, such as the polarization or phase of single photon pulse, the other
is the continuous variable QKD (CV-QKD) where the key information is encoded on continuous Hilbert space, such
as the quadratures of coherent states. Compared with the DV-QKD which is based on single-photon preparation and
detection, CV-QKD can directly utilize the standard telecommunication technologies (such as coherent
detection),  and hence has more potential advantages in practice.

 Generally, the two participants in a CV-QKD
system desire to establish a secret key for a long distance with a
very low signal-to-noise ratio. Then it naturally raises a problem
on designing codes with good error-correction performance, under
such a stringent channel condition. In this case, only low-rate
codes with very long block lengths  can be exploited to achieve
high efficiency key reconciliation.

Low density parity check (LDPC) codes have been shown to possess
Shannon-limit approaching error-correction
performance\cite{748992}, and they have also been broadly applied
in various communication systems, such as the DVB-S2 standard and
the Enhanced Mobile Broadband (eMBB) data channels for 5G New
Radio \cite{ETSI1,Tech}. Chung  et al. designed an irregular LDPC
code of length $10^{7}$ which achieves within 0.04 dB of the
Shannon limit \cite{Chung2001}. Consequently, LDPC codes with long
block lengths have become one of the most promising candidates for
a CV-QKD system. Herein, multi-edge-type (MET) LDPC codes have
attracted much attention due to their excellent  performance.

  The
authors of Ref.\cite{jouguet2012high} reported a random MET-LDPC code with rate 0.1,
where the decoding speed was up to 7.1Mbits/s using a graphic
processing unit (GPU) implementation at signal-to-noise ratio
(SNR) = 0.161, in a CV-QKD system. In Ref.\cite{milicevic2017reconciliation},
the authors employed a quasi-cyclic (QC) MET-LDPC code of
expansion factor 512 and rate 0.1, where the simulation results
demonstrated that the decoding speed was up to 9.17Mbits/s under
the early termination condition at SNR =
0.161. Note that if the messages
of the degree-1 variable nodes are  updated only once at the end
of the iterative decoding procedure, the decoding speed of the
rate-0.1 MET-LDPC code can be up to 30.39Mbits/s when SNR = 0.161
and 64 codewords are decoded simultaneously \cite{Wang2018}.

One of main bottlenecks that restrict the secret key rate of an
LDPC-coded CV-QKD system is the throughput of the
belief-propagation (BP) decoder in the postprocessing of CV-QKD.
This is because successful decoding at very low SNR requires a
large number of iterations. Towards this end, the layered BP
algorithm\cite{layeredBP}  can be utilized to speed up the
decoding convergence. In addition,  a large expansion factor is
employed when constructing QC-MET-LDPC codes in order to making
full use of the parallel computing ability of GPU.

When implementing the GPU-based layered BP decoder, we optimize
the storage of the matrix message by merging bits into one number,
and combine two processes into one kernel to complete a whole
iteration. As a consequence, it reduces the computation amounts.
We also merge the unrelated sub-matrices because they do not
affect each other and can thus be computed simultaneously by
threads. The speed of our layered decoder is up to 64.11Mbits/s
for a code of length $10^6$ and rate 0.1 under the condition of
SNR = 0.161, 50 iterations without early termination.

\section*{Results}


Based on the fact that the LLR messages update at variable/check
nodes can be performed in parallel, the layered BP decoding algorithm is deployed on GPU. This section optimizes the GPU implementation
of layered BP decoding algorithm.

\begin{figure}[ht]
    \centering
    \includegraphics[width=.35\textheight]{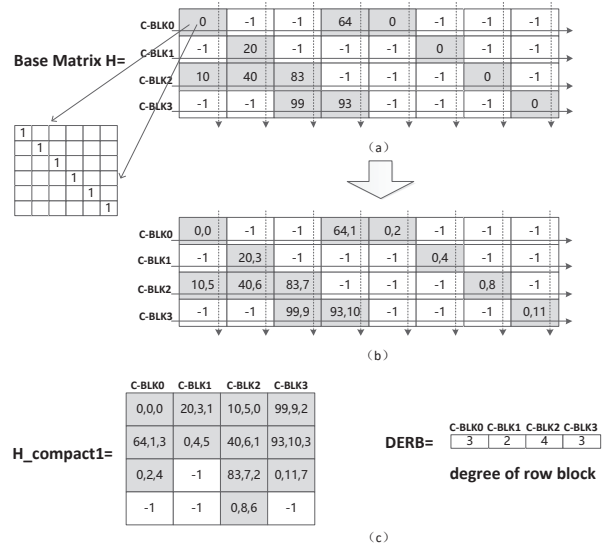}
    \caption{A $4 \times 8$ base matrix and the corresponding file}
    \label{fig6}
\end{figure}

The decoder implementation is  optimized in such a way that the
LLR is stored in global memory for coalesced access. For memory
access in a warp, coalesced access means that the data address of
a thread  always keeps the same  as the thread index, instead of
the unordered access. Since the GPU kernel is executed by a warp
consisting  of $32$ threads, the decoding latency can be hidden
well for a code with length being a multiple of $32$. The layered
BP decoder has a coalesced global memory access, and stores the
parity-check matrix in one file for indexing the corresponding LLR
messages. Such a file denoted by $H\_compact1$, will be applied in
calculating the LLR messages related to the check nodes. Each
element in file $H\_compact1$ contains three pieces of
information:  the amount of the shift, the position of the element
after row rearrangement in the base matrix, and the position of
the column where the non-negative element located in the base
matrix. For example,  Fig. \ref{fig6} displays a $4$-by-$8$ base
matrix with an expansion factor of $100$. Each non-negative
element of the base matrix $\b H$ in Fig.\ref{fig6}-(a) indicates
the amount of shift and  `-1' represents an all-zero matrix. The
second information indicating the position of the element after
row rearrangement is demonstrated in Fig. \ref{fig6}-(b). Then,
one sub-matrices shown in Fig. \ref{fig6}-(c) are used for
indexing the needed messages. Accordingly, the one-dimensional
matrix on the right side in Fig. \ref{fig6}-(c) represent the
degrees of the base matrix (i.e., each element of the the
one-dimensional matrix represents the number of elements non
negative 1 in the corresponding column of the base matrix).

\begin{figure}[ht]
    \centering
    \includegraphics[width=3in]{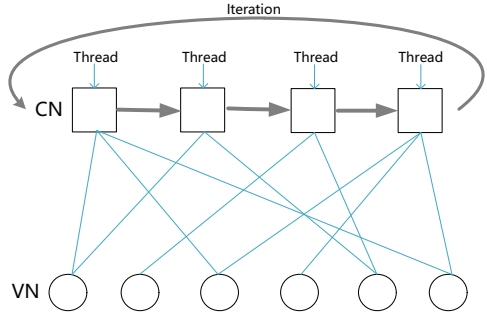}
    \caption{ A node parallel decoding scheme in the layered BP decoder}
    \label{fig8}
\end{figure}

The GPU-based layered BP decoder updates the LLR messages of
variable nodes and check nodes simultaneously, and it also enables
us to decode multiple codewords in parallel. For each individual
codeword, the required number of GPU threads is the same as  the
number of check nodes in a sub-matrix. Each thread computes the
LLR messages received from the neighboring variable nodes, and
also calculates the LLR messages for each adjacent variable node.
This procedure is illustrated in Fig. \ref{fig8} by taking an LDPC
code with $4$ check nodes and $6$ variable nodes as an example. It
is worth noting that at each iteration, one thread corresponds to
a check node. If the expansion factor $Z$ is equal to 100, $1
\times 100$ threads corresponding to check nodes of a sub-matrix
send LLRs to neighboring variable nodes, and also calculate the
messages from variable nodes. Next the $1 \times 100$ reusing
threads update LLRs at the second group of check nodes and their
neighboring variable nodes. The number of reuse of this group of
threads is equal to that of rows of the base matrix. Nonetheless,
the layered BP decoder consumes less thread resources, and the
number of threads for each sub-matrix is only $1 \times 64 \times
Z$ (recall that $Z$ is the expansion factor) when decoding  $64$
codewords simultaneously. The greater the values of $Z$ and the
number of codewords are, the higher the utilization rate of the
thread is.

\begin{figure}[ht]
    \centering
    \includegraphics[width=.35\textheight]{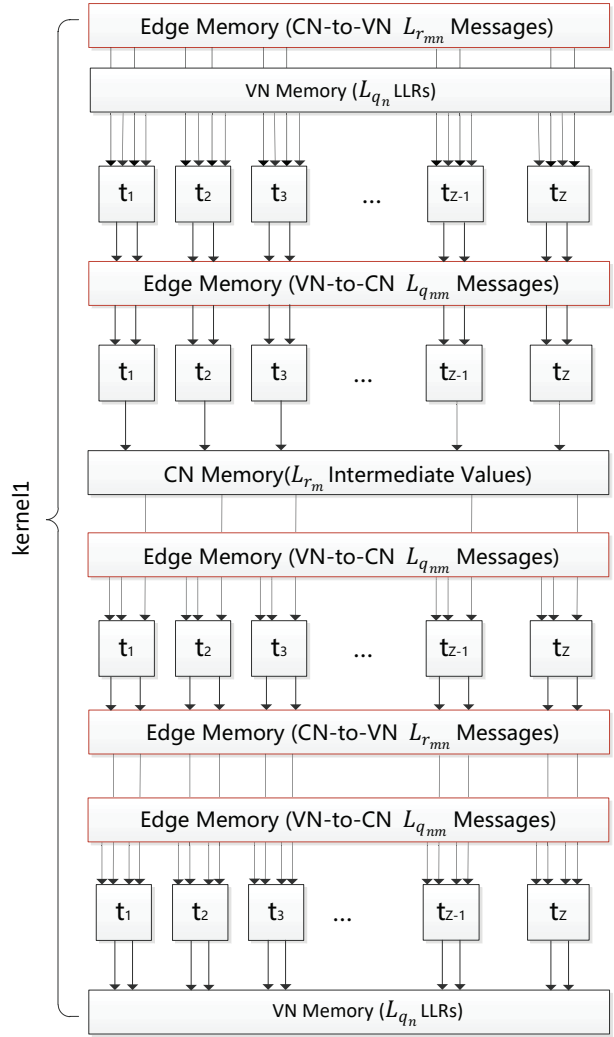}
    \caption{The GPU implementation of the layered decoder showing one multithreaded computation kernel and data flow from top to bottom for one decoding iteration }
    \label{fig9}
\end{figure}

The layered BP decoder only involves  one file $H\_compact1$ for
indexing. This leads to one GPU kernel implementation of the
layered BP decoder for each decoding iteration and each
sub-matrix, whose structure  is demonstrated in Fig.\ref{fig9}. In
the unique kernel, the amount of computation in one thread to
calculate the message from a variable node to a check node or from
a check node to a variable node, defined by the number of edges on
which LLRs are computed, is equal to the degree of the
corresponding check node. The LLR $L^{(t,l)}_{q_{nm}}$ is updated
through $L^{(t-1,l)}_{r_{mn}}$ and $L^{(t,l-1)}_{q_n}$, which
represents the message from the $n$-th variable node to the $m$-th
check node in the $t$-th iteration and the $l$-th layer $l$. Then
each thread calculates the intermediate values $L^{(t,l)}_{r_m}$.
In the remaining part of the kernel, each thread computes the
message from the $m$-th check node to the $n$-th variable node,
denoted by $L_{r_{mn}}^{(t,l)}$, through $L_{r_m}^{(t,l)}$ and
$L_{q_{nm}}^{(t,l)}$. Afterwards the message $L_{q_n}^{(t,l)}$ is
updated through $L_{r_{mn}}^{(t,l)}$ and $L_{q_{nm}}^{(t,l)}$.
There are $Z$ threads in total that are performed at the same
time. The decoder accesses message readily by using $H\_compact1$,
and the LLRs of variable nodes are delivered to the next layer,
i.e., the $(l+1)$-th layer. The above process is a complete
iteration in one layer.

 The original
layered decoder decomposes the $\b H$ matrix into many
sub-matrices on the basis of layers which is equivalent to
treating each layer as a  sub-code.  Each sub-matrix utilizes $1
\times Z$ threads and the serial calculation is conducted among
sub-matrices. In order to increase the layered decoder's thread
utilization, we merge the unrelated sub-matrices into a new
sub-matrix. For example, a $3$-by-$3$ base matrix shown in Eq.
({\ref{m}}) with the expansion factor $Z$ can be divided into
three sub-matrices and the degree of any variable node in each
sub-matrix is equal to one or zero. Herein, a non-negative integer
$a$  in Eq. ({\ref{m}}) such as  `1', `0', and `2'
corresponds to a matrix obtained by cyclically shifting the $Z \times Z$ identity matrix to the right by $a$ bits, and `-1' indicates the all-zero matrix of $Z \times Z$.

\begin{equation}\label{m}
\b H=\quad
\begin{bmatrix}
1&0&-1\\
2&1&1\\
0&2&0
\end{bmatrix}.
\end{equation}

If a base matrix is of the form given in Eq. ({\ref{m1}}), we can
combine its first two rows into one layer; that is, the first two
rows form a sub-matrix in which the degree of any variable node is
equal to one or zero, and the third row separately forms a
sub-matrix. Two sub-matrices wok  in a serial manner by using $2
\times Z$ and $1 \times Z$  threads, respectively.
\begin{equation}\label{m1}
\b H=\quad
\begin{bmatrix}
1&-1&-1\\
-1&2&1\\
2&0&0
\end{bmatrix}.
\end{equation}

Given a base matrix of the form shown in Eq.({\ref{m2}}),  the
first and the third rows of this matrix can be combined into one
sub-matrix, and the second row forms a sub-matrix.
\begin{equation}\label{m2}
\b H=\quad
\begin{bmatrix}
1&-1&-1\\
2&0&0\\
-1&2&1
\end{bmatrix}.
\end{equation}

If all the sub-matrices consist of $K_1$ layers and the number of
codewords that are decoded at the same time is $K_2$ , the thread
utilization rate is computed by $K_1 \times K_2 \times Z \div
67108864$.

Note that too many calls of external functions in a kernel function  spends much time when using CUDA programming. Moreover, the
 warp divergence increases the waiting time when warp threads encounter control flow statements and enter different branches, which
 means that the remaining branches except the branch being executed are blocked at this moment. In this work, the kernel function
  needs to distinguish the sign of the input data, which can be done by calling the application programming interface (API) provided
   by CUDA, thereby avoiding  warp divergence and reducing the calls of external functions. An infinite value or invalid value may
   appear because of the iterative running of the kernel function. To avoid this, an API function for clipping is needed. Another
   optimization is to cut down the branches since the branch structure has great drawbacks when different threads utilize different branches
    with a high  probability. For instance, each thread has different calculation amounts and computation time, and thus the finished threads need
     to wait for other unfinished ones. Towards this end,   we can transform the branch structure to an arithmetic operation when parity checks are used, and hence reduce the decoding time.

\begin{figure}[ht]
    \centering
    \includegraphics[width=.38\textheight]{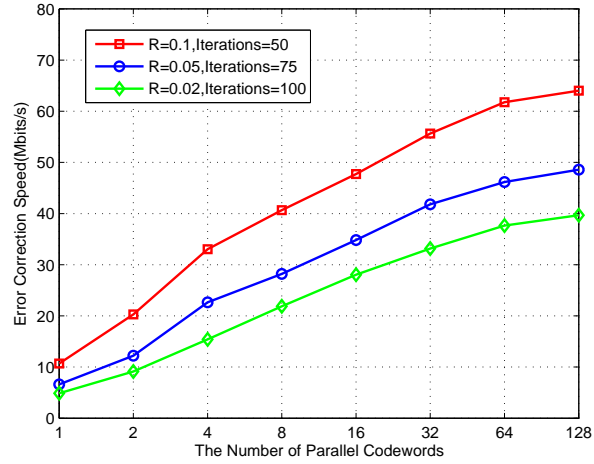}
    \caption{Error correction speed comparison among the different number of decoded codewords in the layered decoder.}
    \label{fig10}
\end{figure}

The performance of GPU-based layered BP decoders are investigated
for QC-MET-LDPC codes with rates $0.1, 0.05$ and $0.02$ on an
NVIDIA TITAN Xp GPU, where the expansion factor is $2500$. Fig.
\ref{fig10} demonstrates the error correction speed when different
number of codewords are decoded simultaneously. The speed grows
steadily from 1 to 128 codewords, and it does not converge even if
the number of codewords reaches 128. Due to the shortage of
storage space, the decoding speed is not considered when the
number of codewords decoded simultaneously exceeds 128. Thus, the
proposed layered BP decoder in this paper  decodes 128 codewords
simultaneously, and its thread utilization rate is $1 \times 128
\times 2500 \div 67108864=0.00477$ where  each sub-matrix consists
of one layer of the base matrix.

Table \ref{tabnoC} compares the performance of the layered BP
decoder with two types of sub-matrices. The one consists of a
single layer and the other consists of multiple layers. When
decoding 128 codewords of lengths $10^6$ simultaneously, the
layered BP decoder with sub-matrices formed by multiple layers
performs better than its counterpart in terms of decoding
throughput. The improvement is 3.2Mbits/s, 1.9Mbits/s and
1.6Mbits/s, respectively, when three rates 0.1, 0.05 and 0.02 are
considered. The method of combining uncorrelated sub-matrices
needs to be further improved, and hence speeds up the decoding and
promotes the thread utilization.

\begin{table}[ht]
    \centering
    \renewcommand\arraystretch{2}
    \begin{tabular}
        {|l|l|l|l|l|l|l|}
        \hline
        Code rate&0.1&0.1&0.05&0.05&0.02&0.02\\
        \hline
        SNR&0.161&0.161&0.076&0.076&0.03&0.03\\
        \hline
        Number of iterations&50&50&75&75&100&100\\
        \hline
        Form of sub-matrix&single&multiple&single&multiple&single&multiple\\
        \hline
        Latency per iteration(s)&0.0164&0.0156&0.0214&0.0206&0.0263&0.0253\\
        \hline
        Error correction speed(Mbits/s)&60.91&64.11&46.72&48.65&37.96&39.51\\
        \hline
    \end{tabular}
\caption{Performance comparison of the layered BP decoder when
decoding different forms of  sub-matrices.}
    \label{tabnoC}
\end{table}

\section*{Discussion}

\begin{table*}[!hbt]
    \centering
        \renewcommand\arraystretch{2}
        \begin{tabular}{|l|l|l|l|}
            \hline
            Code rate&0.1&0.05&0.02\\
            \hline
            SNR&0.161&0.076&0.03\\
            \hline
            $\beta$&92.86\%&94.63\%&93.80\%\\
            \hline
            Number of iterations&50&75&100\\
            \hline
            Total number of edges&3767500&3480000&3337500\\
            \hline
            FER&0.179688&0.25&0.328125\\
            \hline
            Number of codewords&128&128&128\\
            \hline
            Latency per iteration$(s)$&0.0156&0.0206&0.0253\\
            \hline
            Error correction speed $(Mbits/s)$&64.11&48.65&39.51\\
            \hline
    \end{tabular}
    \caption{Performance of the layered BP decoder with GPU implementation.}
    \label{tab1}
\end{table*}
As described in Ref.\cite{milicevic2017reconciliation}, the early termination scheme can
be used as an efficient way to reduce the complexity of LDPC
decoder and avoids unnecessary iterations at high SNR. However,
this scheme may not be efficient at low SNR since the decoding
often fails after multiple iterations in this case. Table
\ref{tab1} illustrates the performance of layered BP decoders
without early termination when three rates $0.1$,$0.05$ and $0.02$
are considered, respectively.  In the decoding process, the
layered BP decoder
 uses the sub-matrices which consist of unrelated layers of the base matrix. In Table \ref{tab1}, the first row represents code rates, and the second one is SNR under the BIAWGNC. The third row $\beta$ indicates the reconciliation efficiency related to the code rate and the number of discarded parity bits, which has an influence on the reconciliation distance and the secret key rate. The fifth row
represents the number of edges of Tanner graph involved in the
decoding, and the eighth row displays the decoding time
consumption within a iteration. Accordingly, the decoding speed  for three tested codes is 64.11Mbits/s, 48.65Mbits/s and 39.51Mbits/s, respectively.

Table \ref{tabearly} demonstrates the performance comparison of
the layered BP decoders with or without early termination at
different SNRs where the code length and the code rate are  $10^6$
and 0.1, respectively. When SNR = 0.161 and 0.171, the layered BP
decoder without early termination performs a little faster than
that with early termination since less calculations are required
to determine whether a valid codeword is obtained in the former.
Nevertheless, When SNR = 0.181,  the layered BP decoder with early
termination is better and the corresponding decoding speed is up
to 93.49Mbits/s. This improvement is attributed to the fact that
the introduction of the early termination condition reduces the
number of iterations significantly at high SNR.

\begin{table}[ht]
    \centering
    \renewcommand\arraystretch{2}
    \begin{tabular} {|l|l|l|l|l|l|l|}
        \hline
        SNR&0.161&0.161&0.171&0.171&0.181&0.181\\
        \hline
        Early termination&No&Yes&No&Yes&No&Yes\\
        \hline
        FER&0.1797&0.1797&0.0273&0.0273&0&0\\
        \hline
        Max iterations&50&50&50&50&50&50\\
        \hline
        Average iterations&50&50&50&50&50&30\\
        \hline
        Latency per iteration(s)&0.0156&0.0162&0.0156&0.0162&0.0156&0.0107\\
        \hline
        Error correction speed(Mbits/s)&64.11&61.74&64.11&61.89&64.11&93.49\\
        \hline
    \end{tabular}
    \caption{Performance comparison of the layered BP decoder when decoding $128$ codewords with/without early termination.}
    \label{tabearly}
\end{table}

We also compare the performance of the layered BP decoder with
other decoders given in previous works, when the code rate is
$0.1$ and SNR = 0.161. As can be seen from Table \ref{tab2}, the
decoding speed of the decoder given in   Ref.\cite{jouguet2012high}  was 7.1Mbits/s with
MET-LDPC codes. In Ref.\cite{milicevic2017reconciliation} , the decoder achieved 9.17Mbits/s
throughput with QC-LDPC codes under the early termination
condition, and the speed reduced to 8.21Mbits/s when executing the
whole iteration process. In Ref.\cite{Wang2018}, the decoding speed of
30.39Mbits/s was obtained by using MET-LDPC codes and decreasing
the computations of variable nodes with degree-$1$. In contrast,
the decoding speed of the proposed layered BP decoder reaches
64.11Mbits/s with no performance degradation by decoding 128
codewords in parallel.

\begin{table}[ht]
    \centering
    \scriptsize
    \renewcommand\arraystretch{2}
    \begin{tabular}{|l|l|l|l|l|l|l|}%
        \hline
        Refs.&Code type&Block length&Average number of iterations&FER&Latency per iteration(s)&Decoding speed (Mbits/s)\\
        \hline
        $[7]$&MET-LDPC&$2^{20}$&100&0.04&1.48&7.1\\
        \hline
        $[8]$&QC-MET-LDPC&$2^{20}$&78&0.0243&1.47&9.17\\
        \hline
        $[9]$&MET-LDPC&$10^6$&100&0.1094&0.0329&30.39\\
        \hline
        This work&QC-MET-LDPC&$10^6$&50&0.1797&0.0156&64.11\\
        \hline
    \end{tabular}
\caption{Performance comparison with prior works using different types of codes, for rate-0.1 and SNR=0.161 in BIAWGNC.}
    \label{tab2}
\end{table}

Despite the high throughput of the GPU-based layered BP decoder,
it also has
 some shortcomings. On one hand,  most of threads are not used and thus
there
exists much space to increase the number of codewords that are decoded simultaneously; but on the other hand,
memory shortage limits the number of parallel decoding. Our future
work will  focus on cutting down on the memory consumption when
decoding one codeword and hence increase the thread utilization by
decoding more codewords simultaneously.

\section*{Methods}

In this paper, the layered BP decoding algorithm is employed, which uses less number of decoding iterations to achieve almost the same performance as the flooding BP algorithm. The procedure of the layered BP decoding algorithm is described as follows.

Let $m$ and $n$  denote the $m$-th check node and $n$-th variable
node, respectively. Moreover, we denote by $L_{q_{nm}}$ the LLR
passed from the $n$-th variable node to $m$-th check node, and by
$L_{r_{mn}}$ the LLR update for the opposite direction. The
superscript $(t)$ indicates the current iteration number, and the maximum number of
decoding iterations is $T$. Let $l$ indicate the current check
node index, which is also  the current layer index of $\b H$. Note
that the total number of layers in $\b H$ is $L$. The layered BP
decoding algorithm views each check node as a sub-matrix, while
the flooding BP decoding algorithm views all check nodes as a
whole matrix.

$step1$: Initialize the LLR value of the variable nodes by using
the received codeword $\boldsymbol{R}$ and channel information
where $\sigma^2$ denotes the variance of noise in BIAWGNC. Let $L^{(t,l)}_{q_n}$ denote the LLR of the $n$-th variable node at the $t$-th iteration and the $l$-th sub-matrix. 
\begin{equation}
L^{(0,0)}_{q_n}= \frac{2R_n}{\sigma^2}
\end{equation}

$step2$: Update the LLR transmitted from check nodes to variable
nodes. Note that $sgn(x)$ is a sign function and $\Phi (x)=
\Phi^{-1} (x) = -\ln(\tanh(x/2 ))$. Let  $N(m)$ represent a set of
variable nodes connected to the $m$-th check node and $N(m)
\backslash n$ excludes the $n$-th variable node from $N(m)$.
Similarly, we use $M(n)$ to denote a set of check nodes connected
to the $n$-th variable node and $M(n) \backslash m$ excludes the
$m$-th check node from $M(n)$. Next, $\boldsymbol{s}$ represents
the syndrome of the received vector where $s_m=0$ and $s_m=1$
indicate the syndrome of the $m$-th check node is even and odd
parity, respectively.
\begin{equation}
sgn(L^{(t,l)}_{r_{mn}}) = \prod_{n^\prime \in N(m) \backslash n} sgn(L_{q_{n^\prime m}}^{(t-1,l)})
\end{equation}
\begin{equation}
|L^{(t,l)}_{r_{mn}}|= \Phi ^{-1}(\sum_{n^\prime \in N(m) \backslash n} \Phi (|L^{(t-1,l)}_{q_{n^\prime m}}|))
\end{equation}
\begin{equation}
L^{(t,l)}_{r_{mn}}=sgn(L^{(t,l)}_{r_{mn}}) \times |L^{(t,l)}_{r_{mn}}|
\end{equation}
\begin{equation}L^{(t,l)}_{r_{mn}}= \begin{cases}
L^{(t,l)}_{r_{mn}} ,\quad s_m=0 \\
-L^{(t,l)}_{r_{mn}},\quad s_m=1
\end{cases}
\end{equation}

$step3$: Update the LLR transmitted from variable nodes to check
nodes.
\begin{equation}
L^{(t,l)}_{q_{nm}}=L^{(t,l-1)}_{q_n}- L^{(t-1,l)}_{r_{mn}}
\end{equation}
\begin{equation}
L^{(t,l)}_{q_n}=L^{(t,l)}_{q_{nm}}+ L^{(t,l)}_{r_{mn}}
\end{equation}

$step4$: Decide the codeword  $\boldsymbol{c}$ by the LLR value of
the variable node.
\begin{equation}
c_n= \begin{cases}
0 ,\quad L^{(t,l)}_{q_n} \ge 0 \\
1,\quad otherwise
\end{cases}
\end{equation}

\bibliography{sample}

\section*{Acknowledgements (not compulsory)}

This work was supported in part by China NSF under Grants
61901425, 61771081, 61771439, 61702469, in part by National
Cryptography Development Fund under Grant MMJJ20170120, in part by
Fundamental Research Funds for the Central Universities under
Grant 2019CDXYJSJ0021, in part by Sichuan Youth Science and
Technology Foundation under Grants 2017JQ0045 and 2019JDJ0060, and
in part by CETC Fund under Grant 6141B08231115.

\section*{Author contributions statement}

Yang Li, Yong Li and B. Xu proposed and guided the work. X. Zhang,
L. Ma, J. Yang and W. Huang designed and performed the experiment.
All authors analyzed the results and wrote the manuscript.

\section*{Additional information}

\textbf{Competing interests:} (The authors declare no competing
interests.

The corresponding author is responsible for submitting a \href{http://www.nature.com/srep/policies/index.html#competing}{competing interests statement} on behalf of all authors of the paper.
 This statement must be included in the submitted article file.

\end{document}